\documentclass[useAMS,usenatbib]{mnras}

\usepackage{newtxtext,newtxmath}
\usepackage[T1]{fontenc}
\usepackage{ae,aecompl}

\usepackage{graphicx}
\usepackage{amsmath}
\usepackage{amssymb}
\usepackage{float}

\graphicspath{{./}{figures/}}

\newcommand{\Msun}{\ensuremath{\mathrm{M}_\odot}}
\newcommand{\ud}{\ensuremath{\mathrm{d}}}

\title[The impact of fallback in CCSNe]{The impact of fallback on the compact remnants and chemical yields of core-collapse supernovae}

\author[C. Chan et al.]{
Conrad Chan$^{1,2}$\thanks{E-mail: conrad.chan@monash.edu},
Bernhard M\"uller$^{1-3}$ and 
Alexander Heger$^{1-5}$
\\
$^{1}$School of Physics and Astronomy, Monash University, VIC 3800, Australia\\
$^2$Joint Institute for Nuclear Astrophysics - Center for the Evolution of the  Elements (JINA-CEE), Monash University, Vic 3800, Australia\\
$^3$OzGrav-Monash -- Monash Centre for Astrophysics, School of Physics and Astronomy, Monash University, VIC 3800, Australia\\
$^4$Center of Excellence for Astrophysics in Three Dimensions (ASTRO-3D), Australia\\
$^5$Tsung-Dao Lee Institute, Shanghai 200240, People's Republic of China
}

\pubyear{2020}

\begin{document}

\label{firstpage}
\pagerange{\pageref{firstpage}--\pageref{lastpage}}
\maketitle

\begin{abstract}
Fallback in core-collapse supernovae plays a crucial role in determining the properties of the compact remnants and of the ejecta composition.  We perform three-dimensional simulations of mixing and fallback for selected non-rotating supernova models to study how explosion energy and asymmetries correlate with the remnant mass, remnant kick, and remnant spin.  We find that the strongest kick and spin is imparted by partial fallback in an asymmetric explosion. Black hole (BH) kicks of several hundred $\mathrm{km}\,\mathrm{s}^{-1}$ and spin parameters of $\mathord{\sim}0.25$ can be obtained in this scenario. If the initial explosion energy barely exceeds the envelope binding energy, stronger fallback results, and the remnant kick and spin remain small. If the explosion energy is high with respect to the envelope binding energy, there is little fallback with a small effect on the remnant kick, but the spin-up by fallback can be substantial. For a non-rotating $12\,\Msun$ progenitor, we find that the neutron star (NS) is spun up to millisecond periods. The high specific angular momentum of the fallback material can also lead to disk formation around black holes. Fallback may thus be a pathway towards millisecond-magnetar or collapsar-type engines for hypernovae and gamma-ray bursts that does not require rapid progenitor rotation. Within our small set of simulations, none reproduced the peculiar layered fallback necessary to explain the metal-rich iron-poor composition of many carbon-enhanced metal-poor (CEMP) stars.  Models with different explosion energy and different realisations of asymmetries may, however, be compatible with CEMP abundance patterns. 
\end{abstract}


\section{Introduction}

Stars with initial masses greater than $8\,\Msun$ will usually end their lives as a core-collapse supernova \citep[CCSN;][]{1966ApJ...143..626C}. Throughout their lifetimes, they produce successively heavier elements via nuclear fusion, deposited in concentric shells. Fusing iron into heavier elements, however, consumes rather than releases energy, so fusion proceeds no further, depositing an iron core at the center of the star. As this iron core exceeds the Chandrasekhar mass, electron degeneracy pressure becomes insufficient to support the core against gravity, and it collapses until nuclear densities are reached. A shock forms, and with the aid of neutrino heating, convection,
and the standing accretion shock instability, it may successfully expand towards the surface, producing an explosion \citep{2007PhR...442...38J,2012ARNPS..62..407J}. Material that does not reach the escape velocity, or that is decelerated by a subsequent reverse shock, will ultimately be accreted onto the central remnant as ``fallback'' \citep{1971ApJ...163..221C,1989ApJ...346..847C}.

Observational studies have found the progenitors of CCSNe are constrained to the mass range $8-18\,\Msun$ \citep{2009MNRAS.395.1409S,2015PASA...32...16S}. The lower limit is in concordance with stellar evolution models which predict that a minimum mass of $7-8\,\Msun$ is necessary for core collapse \citep{2003ApJ...591..288H,2004MNRAS.353...87E}. Whereas stars more massive than $18\,\Msun$ will also experience core collapse, early estimations by \cite{1999ApJ...522..413F} suggest that black holes will form via the subsequent fallback of material (for helium core mass $>8\,\Msun$), or directly from stars more massive than $40\,\Msun$ (helium core mass $>15\,\Msun$). Furthermore, estimations based on the compactness of the stellar core suggest that shock revival may be more difficult in most stars more massive than $18\,\Msun$ \citep[e.g.][]{2015ApJ...801...90P,2016ApJ...818..124E,2016MNRAS.460..742M}. Inferences made using X-ray binary measurements have also shown that black holes may be formed by the collapse of massive stars with no explosion \citep{2003Sci...300.1119M}.

In the intermediate regime where black holes form via fallback rather than directly, it is unclear if the stalled shock can still be revived. In this scenario, black hole formation may be delayed for long enough for neutrino heating to successfully re-energise the shock. Recently, \cite{2017MNRAS.468.4968A} identified a candidate for a failed supernova which produced a weak optical emission following the disappearance of a $25\,\Msun$ progenitor, providing new evidence that the possibility of a weak explosion despite black hole formation has not been ruled out. Additionally, inferences made from abundance determinations of the most metal-poor stars in our galaxy have for some time suggested that the first generation of stars may have exploded weakly, in which iron and other heavy elements closer to the core are part of the strong fallback, and only lighter elements such as carbon and magnesium located closer to the surface are ejected \citep{2003Natur.422..871U,2006NuPhA.777..424N,2014Natur.506..463K,2015ApJ...806L..16B}.

In most models of successful explosions, significant degrees of asymmetry are involved \citep{2012ARNPS..62..407J}, and if these asymmetries persist in the ejecta at late times, then by the argument of momentum and angular momentum conservation, the remnant will be imparted a kick and/or spin by the fallback \citep{2013MNRAS.434.1355J}. Numerous black holes may have been born with substantial kicks, inferred from positions of low-mass X-ray binaries above the galactic plane \citep{2012MNRAS.425.2799R,2017MNRAS.467..298R} and kinematic constraints on individual X-ray binaries \citep{2002ApJ...567..491P,2005ApJ...625..324W,2005ApJ...618..845G,2009ApJ...697.1057F}. Neutron stars too, are observed to travel at several tens to hundreds of km/s in our galaxy, which exceeds the velocities of their progenitor stars \citep{1994Natur.369..127L,2005MNRAS.360..974H}. 

Models accounting for the energy released by fallback accretion have shown that the observed transients could be as bright as superluminous supernovae \citep{2013ApJ...772...30D,2018SSRv..214...59M,2019ApJ...880...21M}. In the case of rapidly rotating progenitors, fallback could provide power to a collapsar engine, leading to gamma-ray bursts \citep{1993ApJ...405..273W,1999ApJ...524..262M}, or at the very least create outflows affecting the subsequent accretion rate, influencing the final remnant properties \citep{2019arXiv190404835B}.

1D hydrodynamical modelling of fallback was performed by \cite{2008ApJ...679..639Z}, from which \cite{2010ApJ...724..341H} calculated chemical yields of ejecta using parameterised models. Given that mixing is an inherently multidimensional process, fallback models with a focus on the effect of mixing via fluid instabilities on ejecta composition have been run in 2D \citep{2009ApJ...693.1780J,2017MNRAS.467.4731C} as well as 3D \citep{2010ApJ...709...11J,2010ApJ...723..353J}. These calculations, however, are initialised from spherically symmetric parameterised piston explosions, thus do not include any mixing due to large-scale asymmetries originating from the explosion mechanism itself. Simulations have used self-consistent core-collapse models to examine the effect of asymmetric ejecta on remnant kick and spin \citep{2013MNRAS.434.1355J,2013A&A...552A.126W}, but the computational cost has restricted these to only the first few seconds following shock revival. To calculate the effect of fallback on the remnant properties requires simulations that extend to at least shock breakout.

In \cite{2018ApJ...852L..19C}, we followed a core-collapse supernova of a $40\,\Msun$ zero-metallicity star \citep{2010ApJ...724..341H} from black hole formation to shock breakout in 3D for the first time using the neutrino hydrodynamics code \textsc{CoCoNuT-FMT} \citep{2015MNRAS.448.2141M} and moving mesh hydrodynamics code \textsc{Arepo} \citep{2010MNRAS.401..791S}. In this simulation, the initial explosion energy was comparable to the binding energy of the star but still managed to eject the hydrogen envelope. The entire He core, however, was accreted onto the black hole, and the only contribution of heavier elements by the ejected material was a small enrichment in C, N, O, and Ca from the hydrogen burning phase. This model was unable to explain the enhanced abundances of heavier elements observed in present-day metal-pool stars. Likewise, the kick and spin of the black hole at shock breakout were both small, due to the complete accretion of hydrodynamic asymmetries, failing to reproduce the high natal kicks of stellar mass black holes inferred by observations.

Motivated by the highly asymmetric structures and large BH kick measured shortly after shock revival, we hypothesised that it is indeed possible to reproduce observed remnant properties if the explosion mechanism has sufficient energy to eject the asymmetric flows. In this study, we test this by following the fallback of an otherwise identical explosion with roughly twice the initial explosion energy to explore the explosion dynamics in the fallback regime more thoroughly. For comparison, we also simulate the explosion and fallback of a $12\,\Msun$ zero-metallicity star, which is expected to form a neutron star. In comparison to the $40\,\Msun$ model, this star is less tightly gravitationally bound, and thus much more likely to explode. We have chosen this case because it will test if asymmetric fallback after the first seconds of
the explosion can still impart significant momentum onto the remnant to affect observed NS velocities.

\section{Numerical Methods}
 We follow the same methodology as in \cite{2018ApJ...852L..19C,2015PASA...32....9F,2016PASA...33...48M}, which we briefly outline here.  We begin by evolving the progenitors using the stellar evolution code \textsc{Kepler} \citep{2010ApJ...724..341H}.  At the onset of  core collapse, the model is mapped into the relativistic neutrino hydrodynamics code \textsc{CoCoNuT-FMT} \citep{2015MNRAS.448.2141M}. To precipitate shock revival in the $40\,\Msun$ models, we artificially increase the strangeness contribution to the axial vector coupling for neutral current neutrino-nucleon scattering to $g_\mathrm{A,s}=-0.2$ as in \citet{2015ApJ...808L..42M}. For the $12\,\Msun$ explosion, we adopt a more physically realistic value of $g_\mathrm{A,s}=-0.05$ \citep[see][]{2016PhRvC..93e2801H}.

Once the shock is successfully revived and a black hole has formed (in the $40\,\Msun$ case) or the rise of the explosion energy has slowed down (in the $12\,\Msun$ case), neutrino effects are negligible, so we map the model into the moving mesh hydrodynamics code \textsc{Arepo} \citep{2010MNRAS.401..791S}, where we benefit from savings in computation time using adaptive time stepping, and improved resolution in simulating mixing instabilities due to the quasi-Lagrangian method \citep{2018ApJ...852L..19C}. It has been show that mapping between the two codes can be performed with good conservation of mass and energy \citep{2018ApJ...852L..19C}. For the $40\,\Msun$ models, the source of neutrino radiation is shut off when the neutron star collapses into a black hole, and any neutrinos still inside the star escape quickly, so the lack of treatment for neutrinos in \textsc{Arepo} is justified. For the $12\,\Msun$ models, which never form a black hole, mapping to \textsc{Arepo} will result in an underestimation of neutrino heating. To mitigate this error, we have kept the simulation in \textsc{CoCoNuT} for as long as possible, limited by computational resources, before mapping to \textsc{Arepo}. This is well beyond the time of shock revival, so any neutrino heating that we have neglected will not drastically alter the energetics of the explosion. In both cases, neutrinos contribute to the gravitational field in \textsc{CoCoNuT}, so we include the gravitational mass of trapped neutrinos in the remnant mass during mapping.

In this study, we run an enhanced energy $40\,\Msun$ model (\texttt{z40p}), two $12\,\Msun$ models (\texttt{z12}, \texttt{z12L}), and include results from \cite{2018ApJ...852L..19C} (\texttt{z40}; Table \ref{tab:models}). For the two $40\,\Msun$ models, we perform the mapping at BH formation, which occurs at $0.5-0.7\,\mathrm{s}$ post-bounce and for the $12\Msun$ models we perform the mapping at $1.8\,\mathrm{s}$ post-bounce, when the accretion has slowed down considerably and explosion energy plateaus. We include \texttt{z12L} case as a demonstration of the numerical resolution required.

\begin{table*}
	\centering
	\caption{Fallback model input parameters. For the BH-forming $40\,\Msun$ models, the accretion radius is set to a multiple of the Schwarzschild radius $R_\mathrm{s}$, whereas for NS-forming $12\,\Msun$ models, the accretion radius is set to a constant.}
	\label{tab:models}
	\begin{tabular}{lcccc}
		\hline
		Name & Progenitor mass & Mass resolution & Accretion radius & Starting time (post bounce)\\
		\hline
		\texttt{z40}  & $40\Msun$ & $1\times10^{27}\,\mathrm{g}$ & $3R_\mathrm{s}$   & $555\,\mathrm{ms}$  \\
		\texttt{z40p} & $40\Msun$ & $5\times10^{26}\,\mathrm{g}$ & $30R_\mathrm{s}$  & $669\,\mathrm{ms}$ \\
		\texttt{z12L} & $12\Msun$ & $5\times10^{26}\,\mathrm{g}$ & $60\,\mathrm{km}$ & $1847\,\mathrm{ms}$ \\
		\texttt{z12}  & $12\Msun$ & $1\times10^{26}\,\mathrm{g}$ & $60\,\mathrm{km}$ & $1847\,\mathrm{ms}$ \\
		\hline
	\end{tabular}
\end{table*}

\begin{figure}
    \includegraphics[width=\columnwidth]{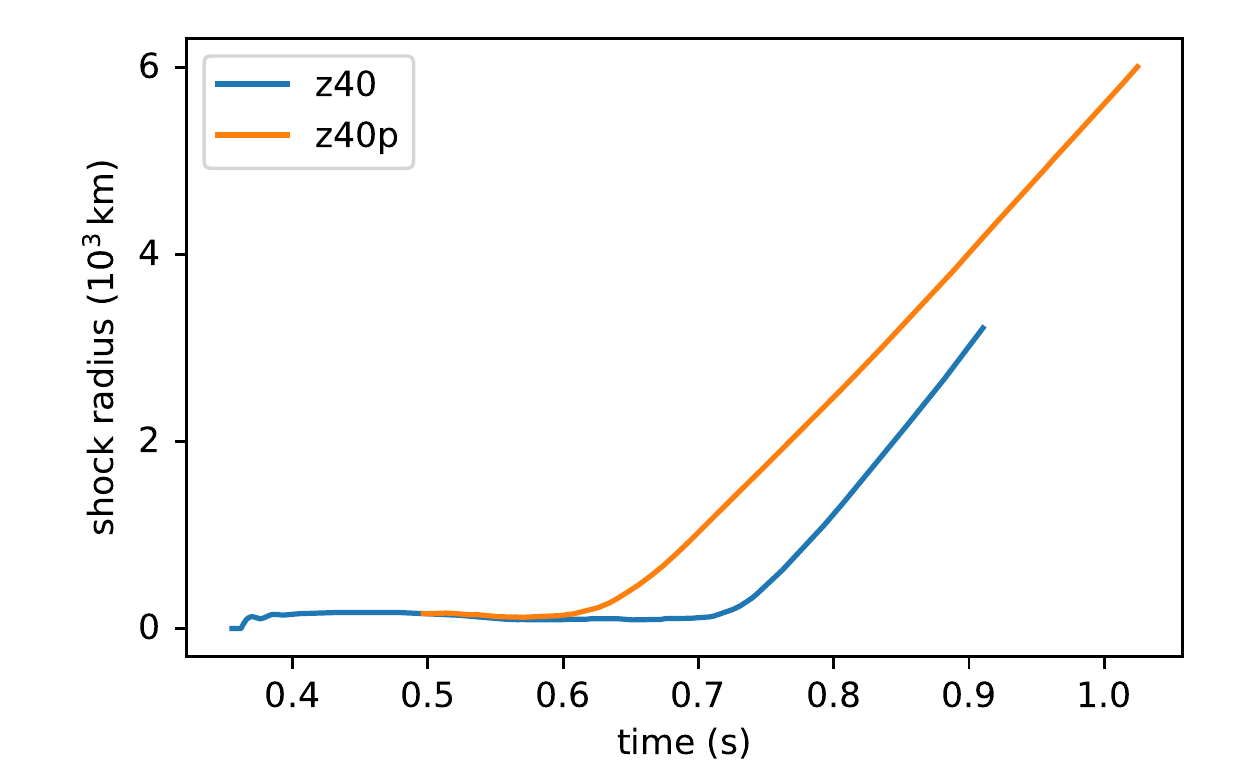}
    \caption{Average shock radius (\textsl{y-axis}) plotted against time (\textsl{x-axis}) since the start of the explosion simulation using \textsc{CoCoNuT-FMT} prior to mapping into \textsc{Arepo}. Density perturbations are introduced at $\mathord{\sim}0.5\,\mathrm{s}$, after which the evolution of the models diverge. Each time-series ends at black hole formation.}
    \label{fig:rshock}
\end{figure}

The \texttt{z40p} explosion is produced by introducing
quadrupolar density perturbations of $\pm 20\%$ ahead of the shock into the density field after collapse. \citet{2016ApJ...833..124M} have shown that pre-shock density perturbations of this scale can arise from oxygen shell burning prior to collapse. The perturbations have the effect of accelerating the growth of instabilities and hence convective energy transport during the explosion phase, leading to earlier shock revival and later black hole formation (Figure \ref{fig:rshock}), resulting in a higher initial explosion energy (Figure \ref{fig:ediag2}). These perturbations are absorbed by convective mixing during the stalled shock phase, and aside from the result of increased explosion energy, do not have a secondary effect on late-stage mixing.

For each of the three fallback simulations, we model accretion onto the central remnant by continuously removing mass inside an accretion radius. This treatment was previously designed to emulate a black hole, which provides no pressure support, but is also a reasonable approximation for neutron stars when material settling on the neutron star cools quickly via neutrinos such that the resulting pressure exerted on infalling gas is negligible. \cite{2018ApJ...852L..19C} shows that the results are relatively insensitive to reasonable changes in the accretion radius, so for \texttt{z40p}, we increase the accretion radius to $30\,R_\mathrm{s}$ to reduce unnecessary computational expense, which allows for a small increase in mass resolution. Here,
\begin{equation}
     R_\mathrm{s} = \frac{2 G M_\mathrm{BH}}{c^2}
\end{equation}
is the Schwarzschild radius, where $M_\mathrm{BH}$ is the instantaneous BH mass.

We have run the $12\,\Msun$ explosion with the same resolution of $5\times10^{26}\,\mathrm{g}$ (\texttt{z12L}), but found that this is insufficient to spatially resolve structures in the flow immediately after mapping. Thus we have increased the resolution to $10^{26}\,\mathrm{g}$ (\texttt{z12}). Since \texttt{z12L} was not advanced any further once this problem was identified, from hereon, we refer only to the more accurate \texttt{z12} model when discussing $12\,\Msun$ explosions. A detailed resolution study is computationally prohibitive, but for the intent of this study, our simulations, which conservatively capture the overall energetics of the system, are sufficient to provide a qualitative understanding of the effect of fallback, especially in light of the many uncertainties present in the modelling parameters.

\section{Results}

\subsection{Explosion Energy}
\begin{figure}
    \includegraphics[width=\columnwidth]{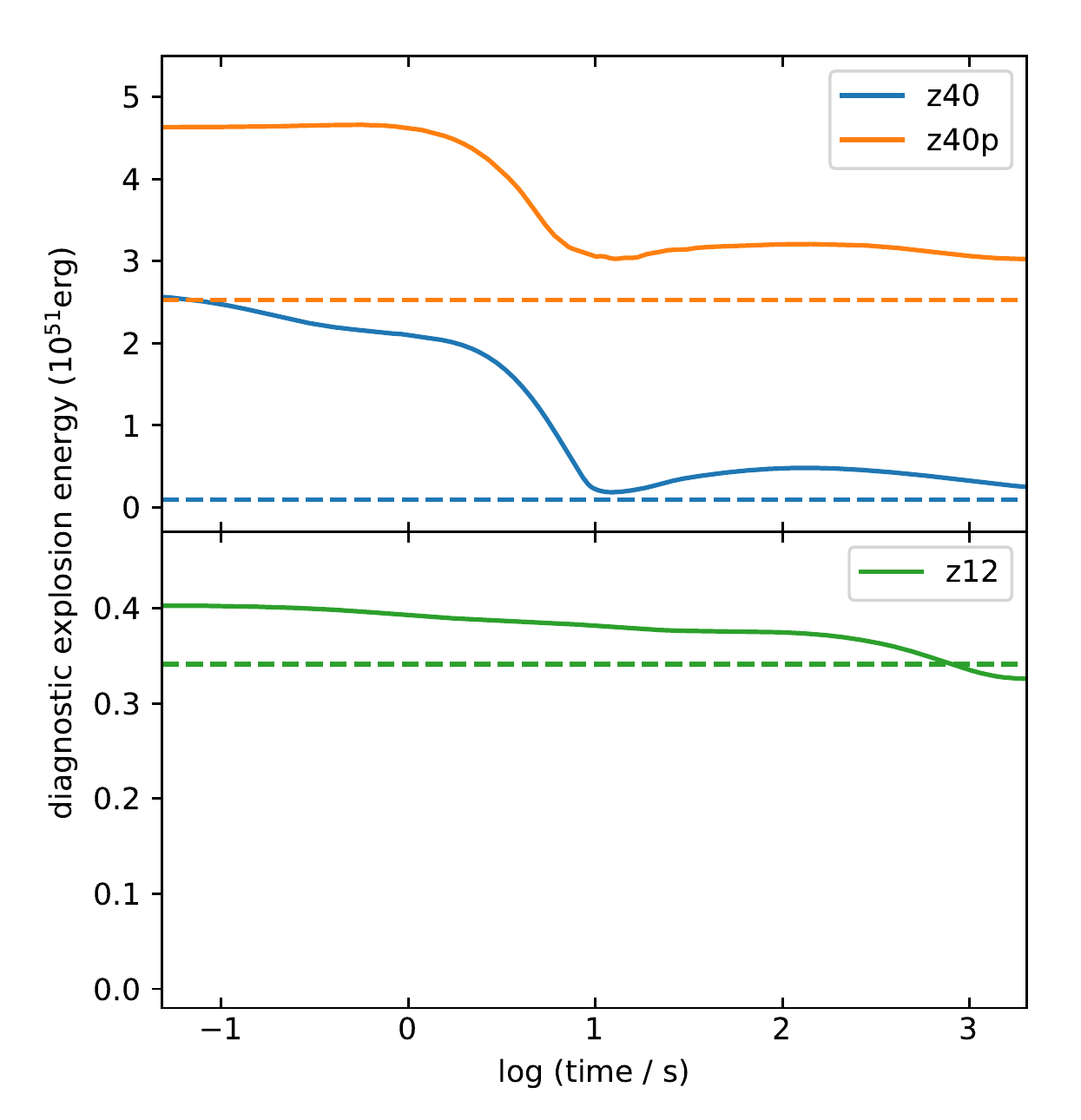}
    \caption{Diagnostic explosion energy (\textsl{y-axis}) as a function of time post-bounce (\textsl{x-axis}). \textsl{Dashed lines}: initial gravitational binding energy of material ahead of the shock subtracted from the initial diagnostic explosion energy, which is a rough prediction of the final explosion energy.}
    \label{fig:ediag2}
\end{figure}

Given the difficulties in predicting the final explosion energy, to monitor the energetics of the explosion as the shock propagates through the envelope, we adopt the commonly used diagnostic explosion energy, $E_\mathrm{diag}$ \citep{2006A&A...457..281B,2017MNRAS.472..491M},
\begin{equation}
    E_\mathrm{diag} = \int_{e_\mathrm{tot}>0} \!\!\!\!\rho\, e_\mathrm{tot}\, \ud V,
\end{equation}
the sum of the total energy (internal, kinetic, and gravitational), $e_\mathrm{tot}$, of all unbound fluid at any point in time (Figure \ref{fig:ediag2}). We find that the final explosion energy of each model is lower than than initial explosion energy. As the hot, rapidly expanding shock moves outwards, it transfers energy into the initially bound (i.e., negative $e_\mathrm{tot}$) envelope. For the envelope to be ejected, it must absorb the corresponding amount of energy from the shock to raise its total energy to greater than zero. Thus we expect the decrease in explosion energy from shock revival up until shock breakout to correspond to the binding energy of the envelope. The material that is initially unbound at shock revival is then depleted of its kinetic energy, and proceeds along a trajectory back towards the central remnant. Indeed, in all three explosions, the difference between the initial diagnostic explosion energy and the binding energy of material ahead of the shock provides a good approximation for the final explosion energy. Pinpointing the exact final explosion energy requires the simulation to be continued beyond breakout, but doing so requires significant computational resources. The diagnostic explosion energy at shock breakout, which has since levelled off, provides an adequate estimate.

\subsection{Remnant properties}

\begin{figure}
    \includegraphics[width=\columnwidth]{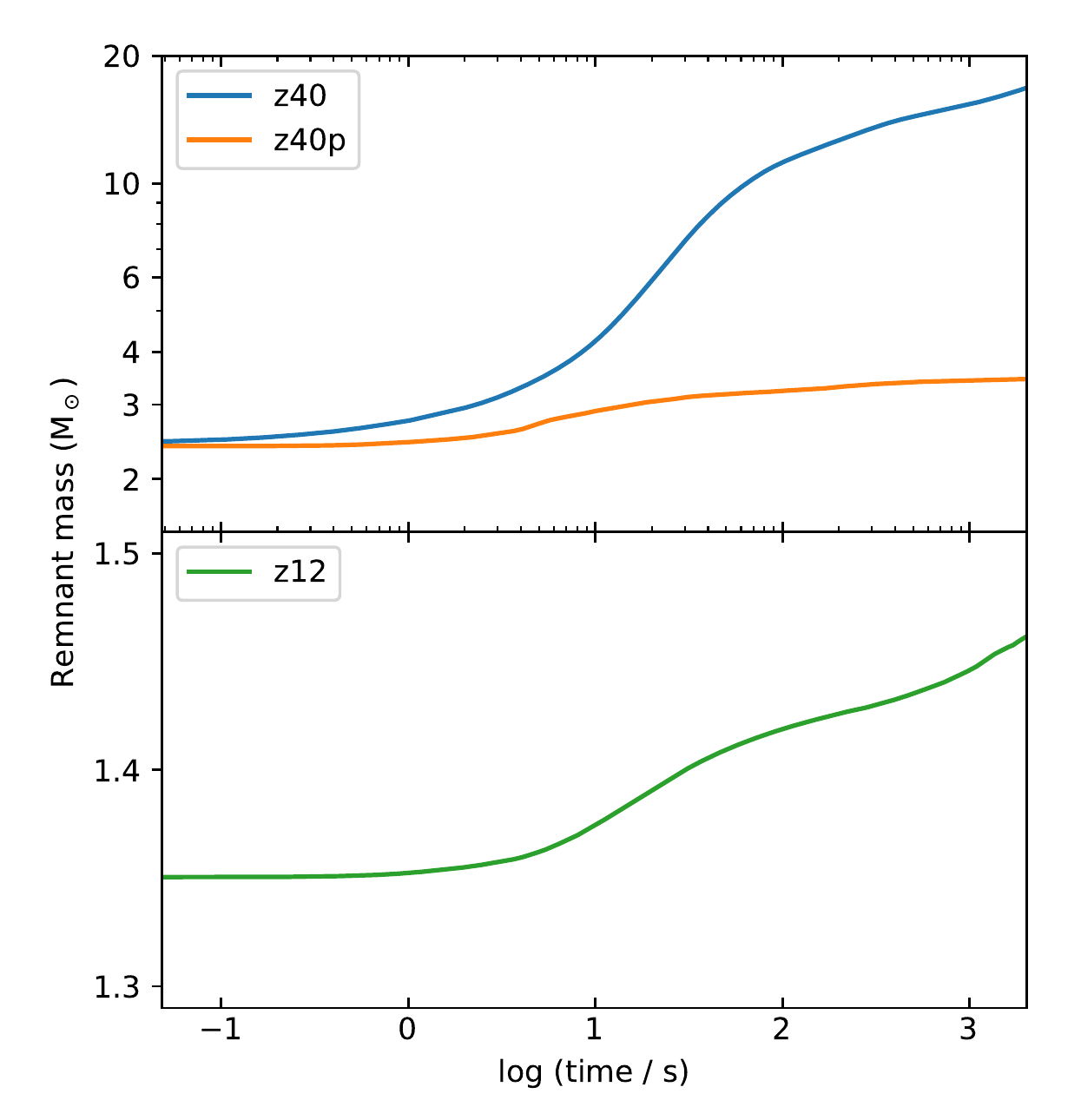}
    \caption{Mass of the central remnant (\textsl{y-axis}) as a function of time post bounce (\textsl{x-axis}). In \texttt{z40} and \texttt{z40p}, the central remnants are black holes, and in \texttt{z12}, the central remnant is a neutron star.}
    \label{fig:bh_mass}
\end{figure}

\begin{figure}
    \includegraphics[width=\columnwidth]{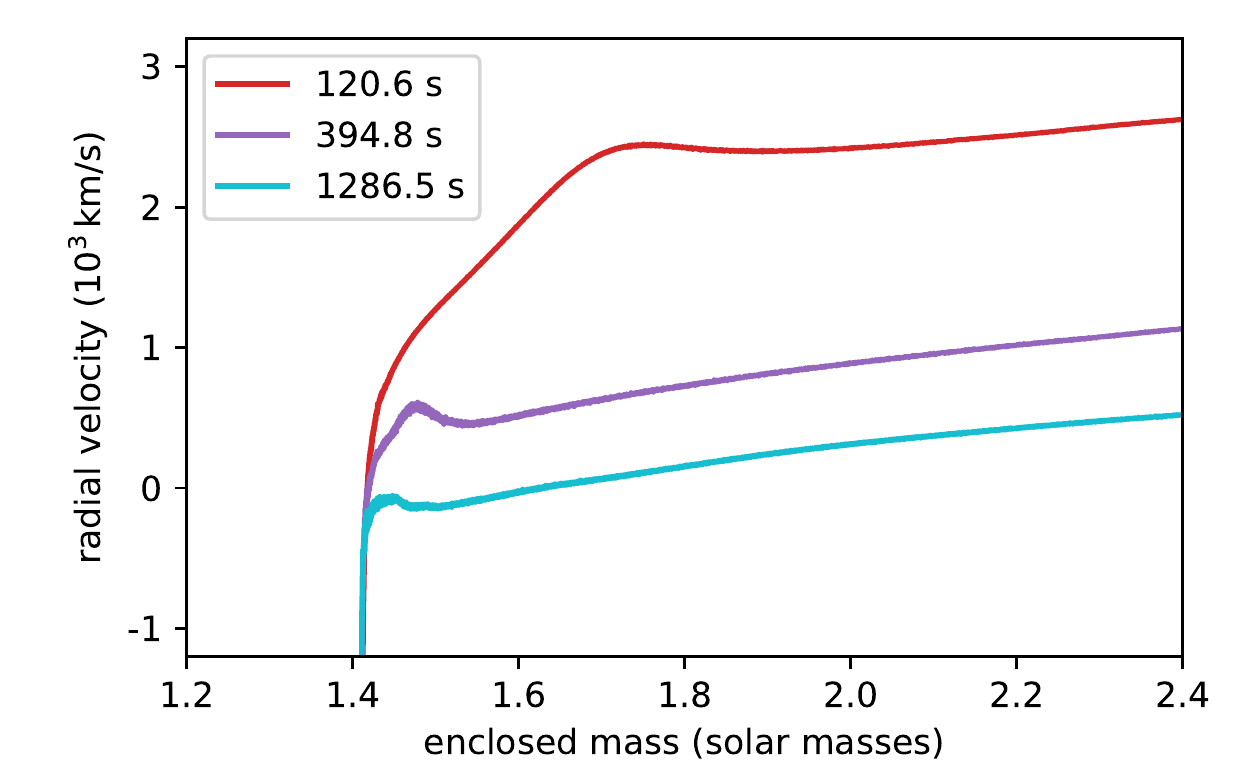}
    \includegraphics[width=\columnwidth]{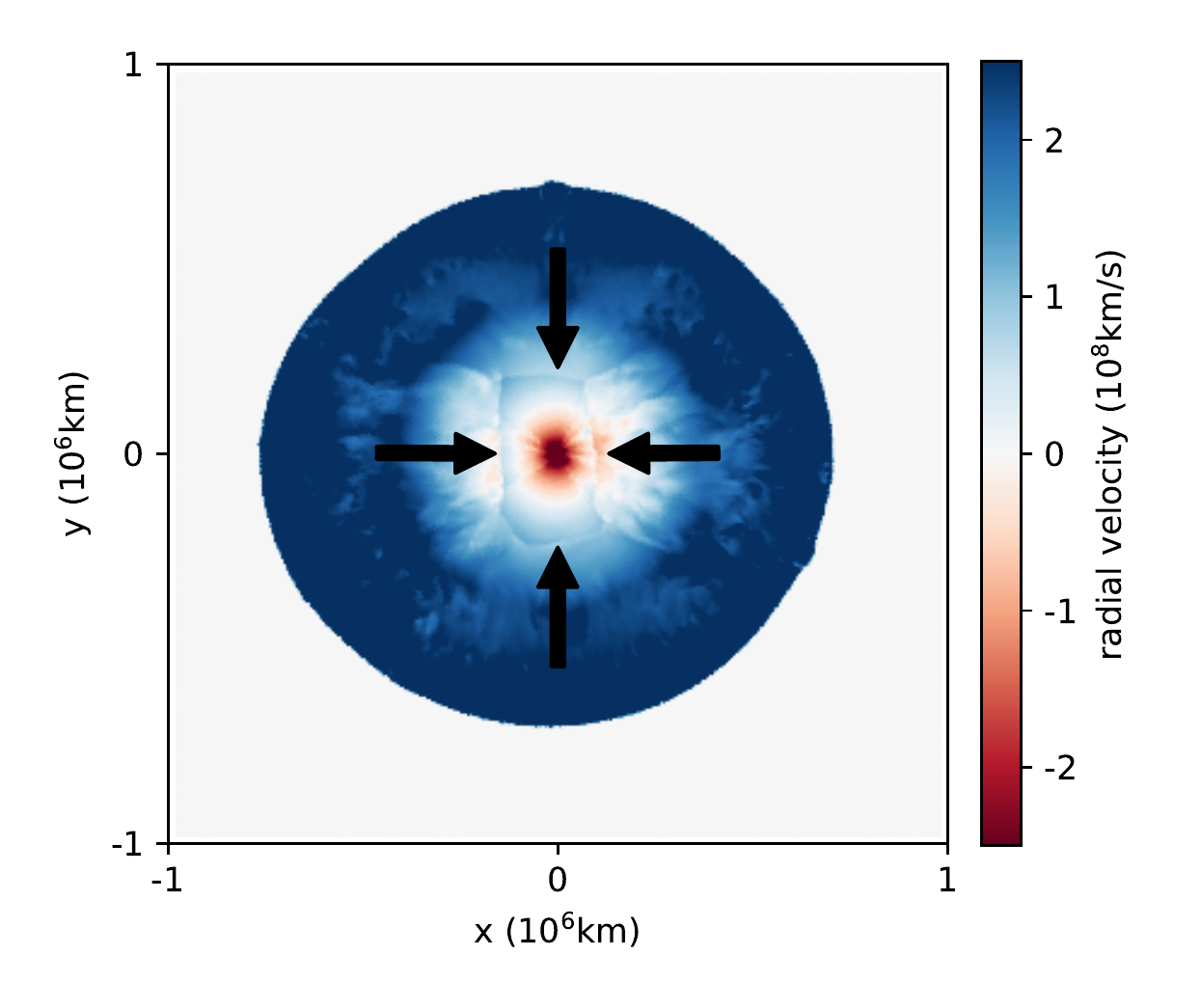}
    \caption{The reverse shock in the \texttt{z12} explosion. \textsl{Upper Panel}: Average radial velocity (\textsl{y-axis}) as a function of mass-coordinate (\textsl{x-axis}) for three points in time. A bump in velocity, the reverse shock, is seen propagating inwards. \textsl{Lower Panel}: 2D slice of radial velocity at $120.5\,\mathrm{s}$ showing the asymmetric structure of the reverse shock. The position of the reverse shock is indicated by arrows.}
    \label{fig:reverse_shock}
\end{figure}

\begin{figure}
    \includegraphics[width=\columnwidth]{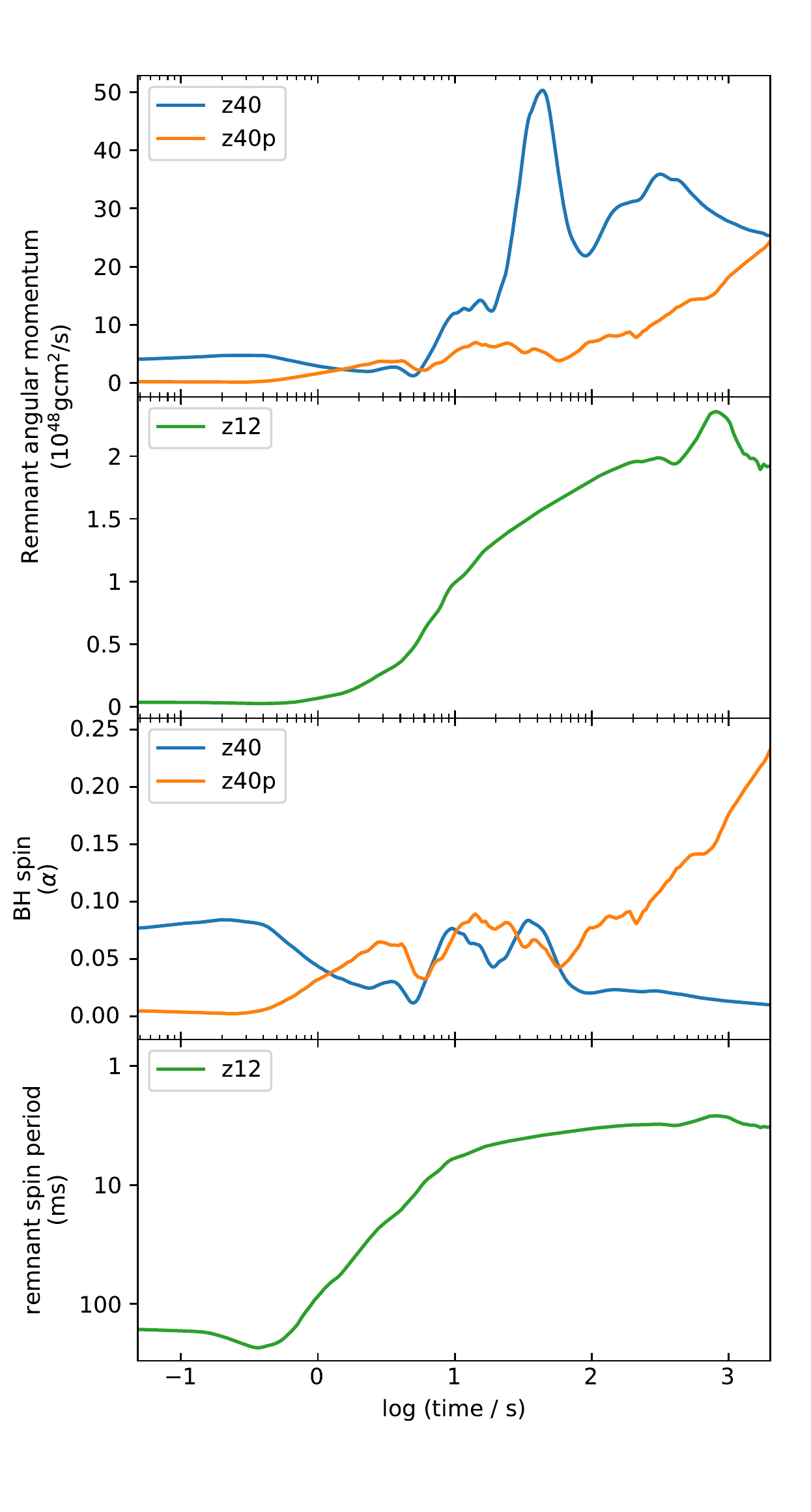}
    \caption{\textsl{Upper Panels}: angular momentum of the remnant (\textsl{y-axis}) as a function of time (\textsl{x-axis}). \textsl{Lower Panels}: for the BH-forming explosions \texttt{z40} and \texttt{z40p}, we plot the spin parameter, and for the NS-forming explosion \texttt{z12}, we plot the NS spin period, which is inversely proportional to the angular velocity.}
    \label{fig:bh_spin}
\end{figure}

\begin{figure}
    \includegraphics[width=\columnwidth]{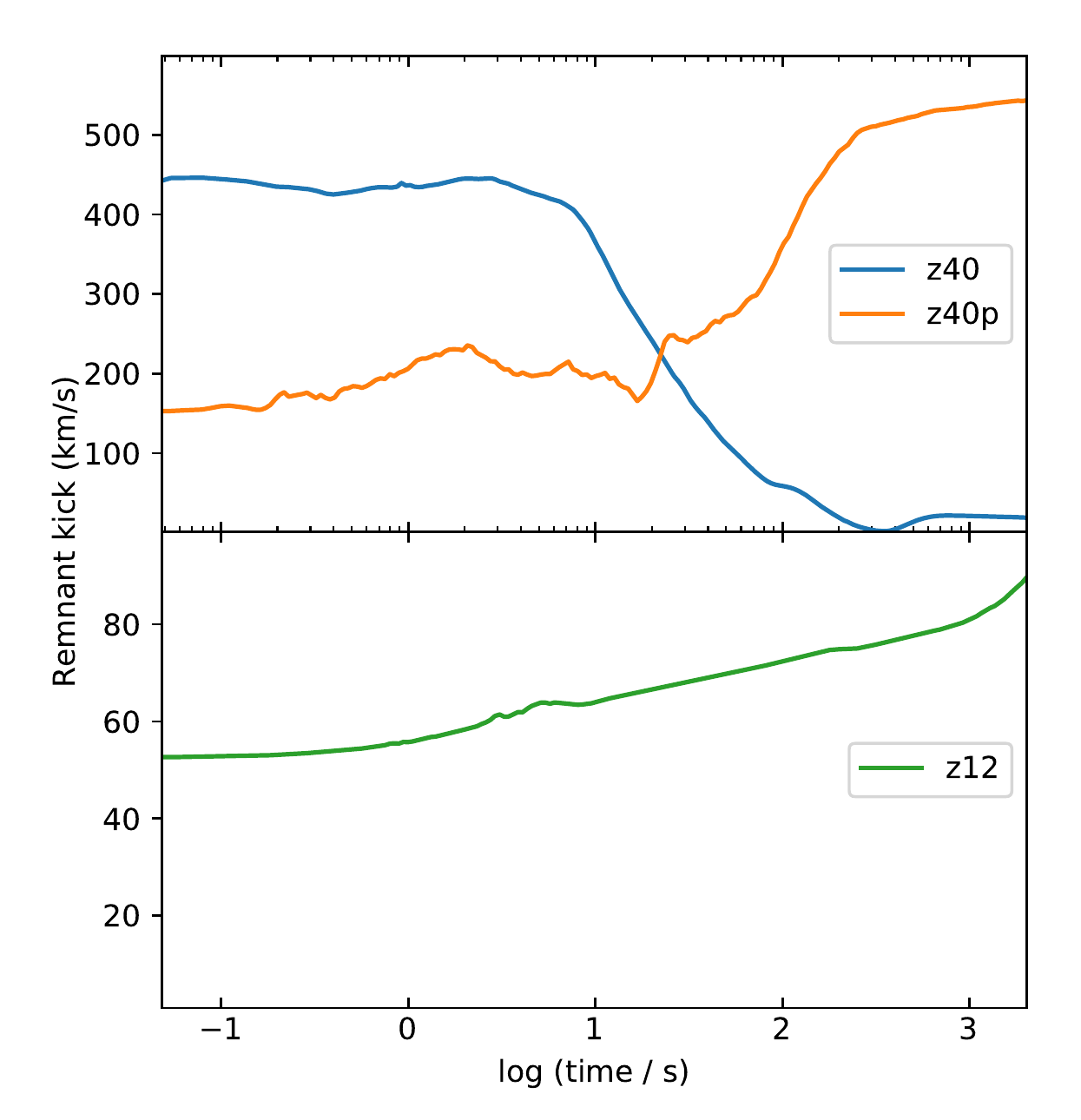}
    \caption{Kick velocity of the central remnant (\textsl{y-axis}) as a function of time (\textsl{x-axis}), calculated using the momentum of the ejecta.}
    \label{fig:bh_kick}
\end{figure}

In Figure \ref{fig:bh_mass}, we show the masses of the central remnants, which are calculated by tracking the accreted mass. In Figures \ref{fig:bh_spin} and \ref{fig:bh_kick}, we show the spins and kicks of the central remnants, which are calculated using global momentum and angular momentum conservation. The accretion rate of the remnant in \texttt{z12} rises at the end of the simulation. This is caused by the formation of a reverse shock (Figure \ref{fig:reverse_shock}) as the forward shock travels into the hydrogen shell. The reverse shock propagates inward, decelerating the post-shock matter and producing a rise in remnant mass close to the time of shock breakout, which can be seen as the drop in explosion energy (Figure \ref{fig:ediag2}). 2D slices of the velocity field reveal that the reverse shock is oblate. As the innermost part of the material that has undergone the reverse shock begins accreting, the remnant angular momentum sharply rises at first, but drops to its original value once the remainder of the reverse shocked material is accreted (Figure \ref{fig:bh_spin}). The kick, on the other hand, is increased by the accretion of the shock (Figure \ref{fig:bh_kick}). 

Asymmetric accretion at late times could lead to asymmetric neutrino emission. In principle, this could lead to further asymmetric flows and enhanced mixing over the first seconds of the explosion. However, previous studies \citep[e.g.][]{2017MNRAS.472..491M,2018ApJ...865...61G} have shown that the neutrino-induced kick reaches a few tens of $\mathrm{km}/\mathrm{s}$ at best, so it would be a subdominant contribution for the black hole formed in \texttt{z40p}.

Although the remnant mass continues to increase at the end of the simulation, subsequent accretion may again be slowed down by the formation of a
negative pressure gradient behind
the reverse shock and an outward-propagating wave \citep{2016ApJ...821...69E}. In \cite{2018ApJ...852L..19C}, we spherically averaged the model to determine the late-time fallback since the shock had already sphericised at shock breakout, but this approach is not possible for the explosions in this study without also discarding the asymmetries present. 

An estimate by summing over the remaining bound mass (Table \ref{tab:ejecta}) suggests that the neutron star in \texttt{z12} will only accrete an additional $0.18\,\Msun$, for a total of $1.64\,\Msun$, well below modern estimates of the limit for collapse to a black hole \citep{2016ARA&A..54..401O}. Interestingly, the expected remnant mass of \texttt{z40p} is $3.82\,\Msun$, which lies within the supposed ``mass gap'' of $3-5\,\Msun$ in compact objects suggested by X-ray binary observations \citep{2010ApJ...725.1918O,2011ApJ...741..103F},  gravitational wave determination of black hole masses \citep{2016PhRvX...6d1015A,2016PhRvL.116x1103A,2017PhRvL.118v1101A} and neutron star masses \citep{2017PhRvL.119p1101A,2020arXiv200101761T}. Recent microlensing estimates, however, have suggested that black holes do indeed exist in this range \citep{2019arXiv190407789W}, and black hole growth via the fallback mechanism that we observe in our models may be necessary to explain their existence
\citep[cp.\ also][for the
effect of fallback in 1D]{2020ApJ...890...51E}.

\begin{table}
	\centering
	\caption{Final remnant and ejecta mass, predicted using the total energy of each mass element.}
	\label{tab:ejecta}
	\begin{tabular}{lcc}
		\hline
		Name & Predicted remnant mass & Predicted ejecta mass\\
		\hline
		\texttt{z40}  & $28.5\,\Msun$ & $11.5\,\Msun$ \\
		\texttt{z40p} & $3.82\,\Msun$ & $36.18\,\Msun$ \\
		\texttt{z12}  & $1.64\,\Msun$ & $10.38\,\Msun$ \\
		\hline
	\end{tabular}
\end{table}

In \texttt{z40p}, the higher explosion energy results in dramatically slower accretion onto the central remnant compared to \texttt{z40}, and a high fraction of escaping asymmetries results in a greater kick and spin (Figure \ref{fig:bh_kick}). Due to the lower explosion energy, the remnant in \texttt{z40} receives a high initial kick and spin, but flattens out to much smaller values as the shock becomes more spherical. In \texttt{z40p} and \texttt{z12}, the remnant properties have not converged at shock breakout, but provide an indication of the kick and spin that can be attained if there is no significant subsequent fallback.

\begin{figure*}
    \includegraphics[width=0.8\textwidth]{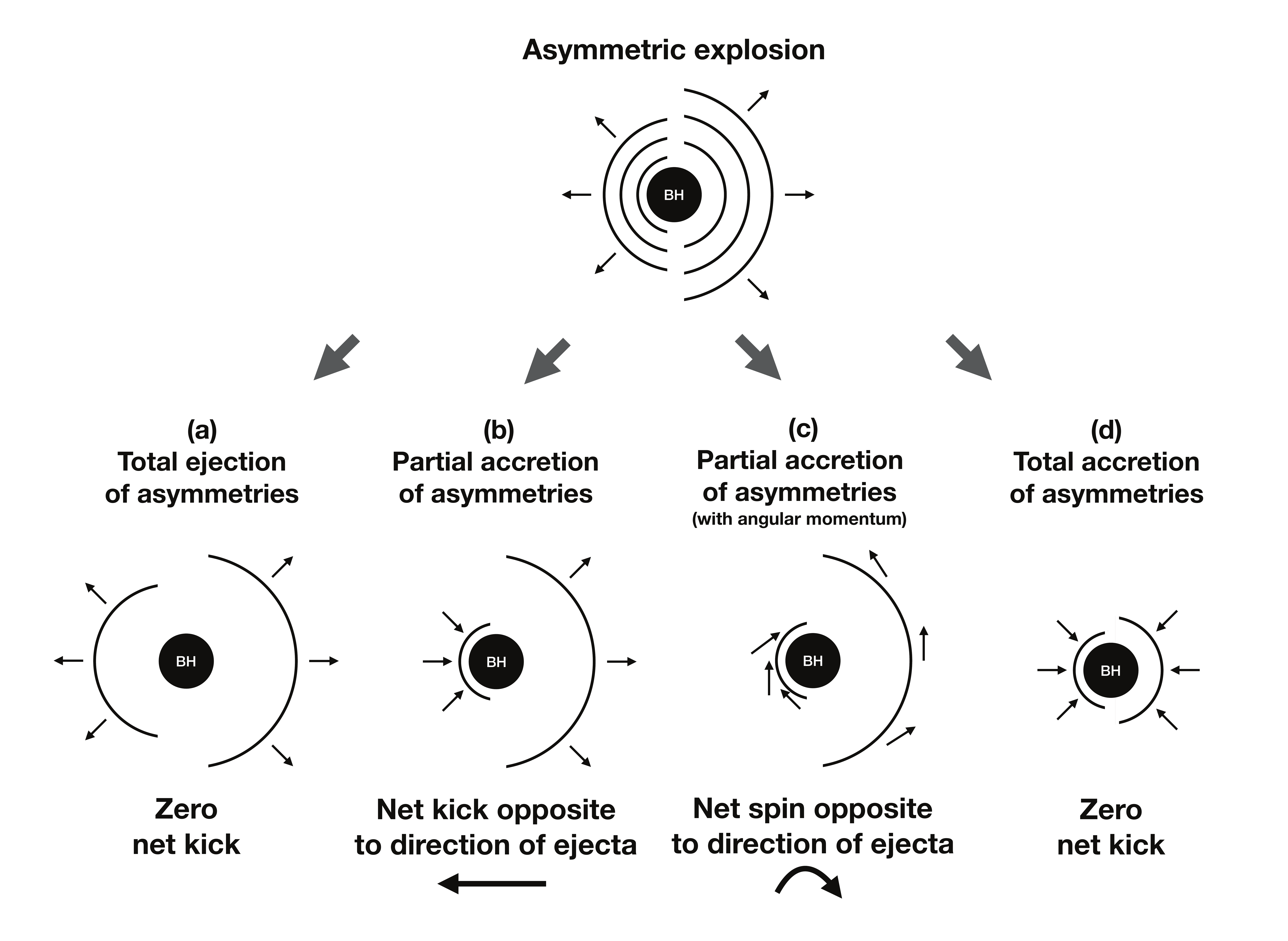}
    \caption{Illustration of how a remnant kick arises from asymmetric fallback. Following an asymmetric explosion, there are three possible scenarios. In the first scenario (a), all of the asymmetries are ejected, and there is zero net kick on the remnant. In the second scenario (b), there is partial accretion of the asymmetries with linear momentum, leading to a net kick on the remnant by argument of momentum conservation. In the third scenario (c), there is partial accretion of the asymmetries with angular momentum, leading to a net spin on the remnant by argument of angular momentum conservation. In the fourth scenario (d), there is total accretion of the asymmetries, and there is zero net kick on the remnant. \texttt{z40} undergoes scenario (a), whereas \texttt{z40p} undergoes a combination of scenarios (b) and (c).}
    \label{fig:fallback_diagram}
\end{figure*}

To explain our findings, we follow the same arguments used in \cite{2017ApJ...837...84J} to estimate the remnant kick velocity. The momentum imparted onto the central remnant is equal and opposite to the momentum of the ejecta,
\begin{equation}
    \left|\textbf{p}_\mathrm{rem}\right| = \left|\textbf{p}_\mathrm{ej}\right| = \sqrt{\alpha\, E_\mathrm{kin}\, M_\mathrm{ej}},
\end{equation}
where
\begin{equation}
    \alpha = \frac{\left|\int_{M_\mathrm{ej}} \rho\,\mathbf{v}\, \ud V\right|}{\int_{M_\mathrm{ej}} \rho\left|\mathbf{v}\right| \ud V}
\end{equation}
is the momentum-asymmetry parameter, $E_\mathrm{kin}$ is the kinetic energy of the ejecta, and $M_\mathrm{ej}$ is the mass of the ejecta. The parameter $\alpha$ is zero for a spherically symmetric ejection, and approaches unity for one-sided ejecta.
It follows that the remnant velocity
$v_\mathrm{rem}$ scales as
\begin{equation}
    v_\mathrm{rem} \sim \frac{\sqrt{\alpha\, E_\mathrm{kin}\, M_\mathrm{ej}}}{M_\mathrm{rem}}.
    \label{eq:vkick}
\end{equation}
The dominant mechanisms for transfer of momentum to the central remnant are the accretion of infalling matter, pressure exerted by surrounding matter, and gravitational acceleration by surrounding matter \citep[``gravitational tug-boat'' mechanism;][]{2013A&A...552A.126W}. The pressure forces can be appreciable depending on the analysis volume
for which the hydrodynamical fluxes and gravitational forces are computed, as shown, for example, by \citet{2013A&A...552A.126W}. After the central remnant collapses into a black hole, it ceases to provide pressure support, so it also ceases to receive a kick contribution via pressure. Although we track all three contributions to the kick in \textsc{CoCoNuT}, our assumption in \textsc{Arepo} that matter accretes supersonically and then cools rapidly as it settles onto the neutron star implies that the pressure contribution is zero. The net momentum of the remnant and any stationary mass element outside of the remnant prior to the explosion remains zero throughout the subsequent evolution. When this mass element is far from the remnant, it exerts a gravitational pull on the remnant, accelerating it towards the mass element. As it is accreted onto the remnant,
there is a an advective flux of momentum
in the opposite direction
onto the remnant , so that the total acceleration is zero. After the remnant has finished accreting, the gravitational tug-boat acceleration caused by ejected matter can continue to accelerate it. It is the gravitational acceleration arising from this escaping matter -- without an advective
momentum flux counterpart -- which imparts a net kick onto the remnant. We illustrate this concept in Figure \ref{fig:fallback_diagram}. Small anisotropies in angular momentum of material at large radii translate to rapid rotational velocities when the material falls inwards. Although the star is spherically symmetric prior to the explosion, mixing induced by the shock imparts local angular momentum to the fluid, even though the total angular momentum necessarily remains zero.

Material that is only weakly accelerated away from the core by the explosion will fall back at early times, whereas matter accelerated more strongly without attaining escape velocity will fall back at later times. Fallback at late times is associated with the accretion of the asymmetric ``counterparts'' of early fallback. A strong fallback at late times results from a weak explosion energy, and is likely to be positively correlated with a small value of $\alpha$. If material is accelerated strongly enough to attain escape velocity by means of a high explosion energy, then the accretion rate at late times will be small, but the resultant kick and spin correspondingly large. Thus we argue if fallback is the main contribution to remnant kick and spin, we expect to find that smaller remnants are born with faster kicks, and vice-versa. We may also expect the spin and kick would typically be positively correlated.

Since the kick velocity is positively correlated with the explosion energy (Equation \ref{eq:vkick}), and the explosion energy most likely increases with progenitor mass \citep{2013MNRAS.436.3224P,2014AstL...40..291C,2019MNRAS.489..641M}, we expect the kick to be positively correlated with remnant mass for the lower mass range
\citep{2016MNRAS.461.3747B,2017ApJ...837...84J,2018MNRAS.481.4009V}. At higher initial masses, however, the core mass does not increase proportionally, and since only $\mathord{\sim} 1\, \Msun$ can be accreted onto the proto-neutron star after the onset of the explosion, 
the amount of energy that can be pumped into the explosion by neutrino heating is limited.
Thus the initial explosion energy will no longer
increase significantly with progenitor mass
for high-mass progenitors.
Instead, the binding energy of the additional mass in the envelope eventually saps a roughly constant initial explosion energy, leading to a weaker final explosion energy, fewer asymmetries, and thus a smaller kick. A more detailed exploration of the parameter space is necessary to confirm this tentative result.

\subsection{Implications for hypernovae and GRBs}
\begin{figure}
    \includegraphics[width=\columnwidth]{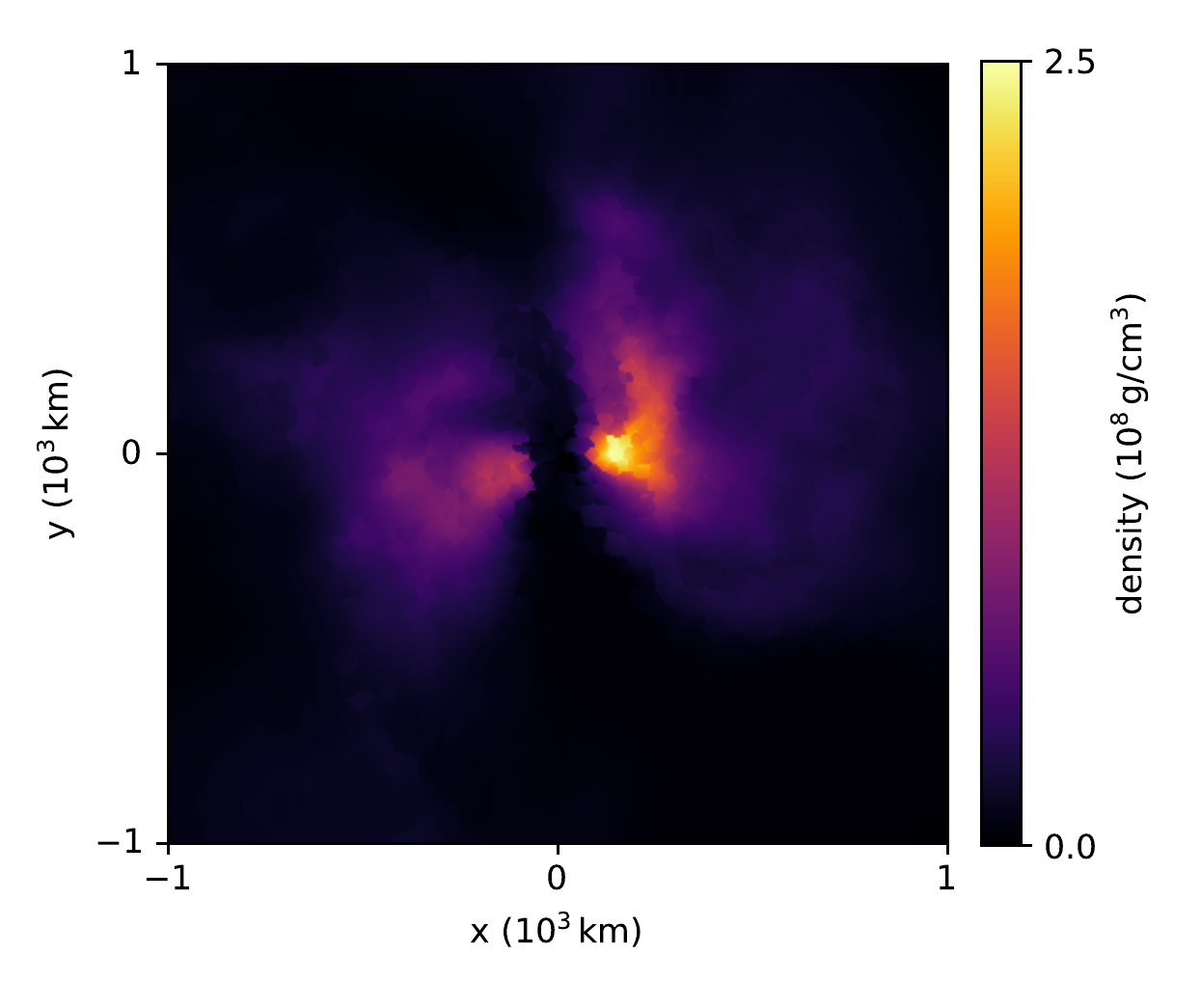}
    \includegraphics[width=\columnwidth]{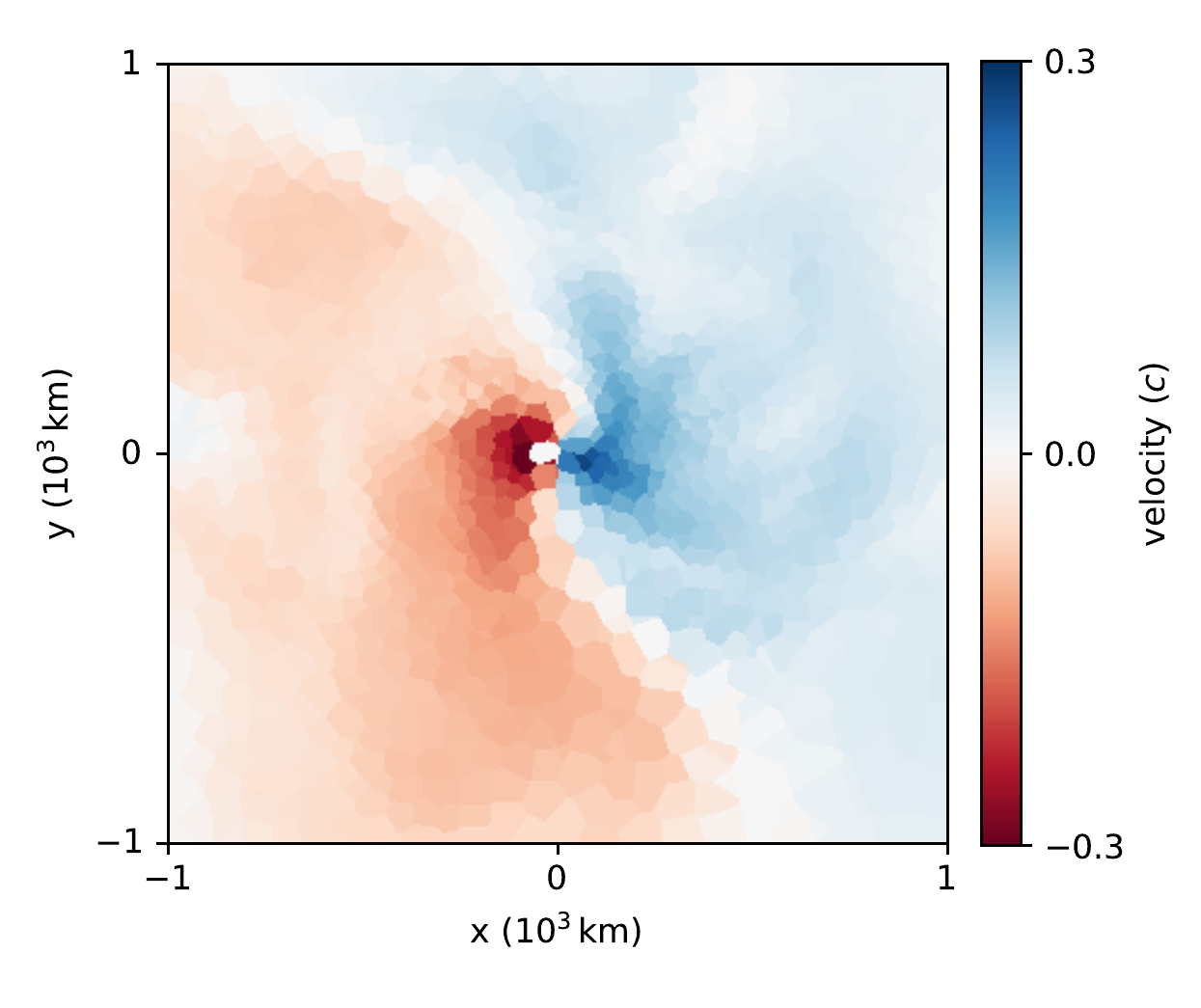}
    \caption{Cross-sections of \texttt{z40p} at $34\,\mathrm{s}$ after bounce showing density (\textsl{Upper Panel}) and velocity in the z-direction (\textsl{Lower Panel}). These point towards disk formation at radii of several $100\,\mathrm{km}$ around black hole.}
    \label{fig:disk_slice}
\end{figure}

\begin{figure}
    \includegraphics[width=\columnwidth]{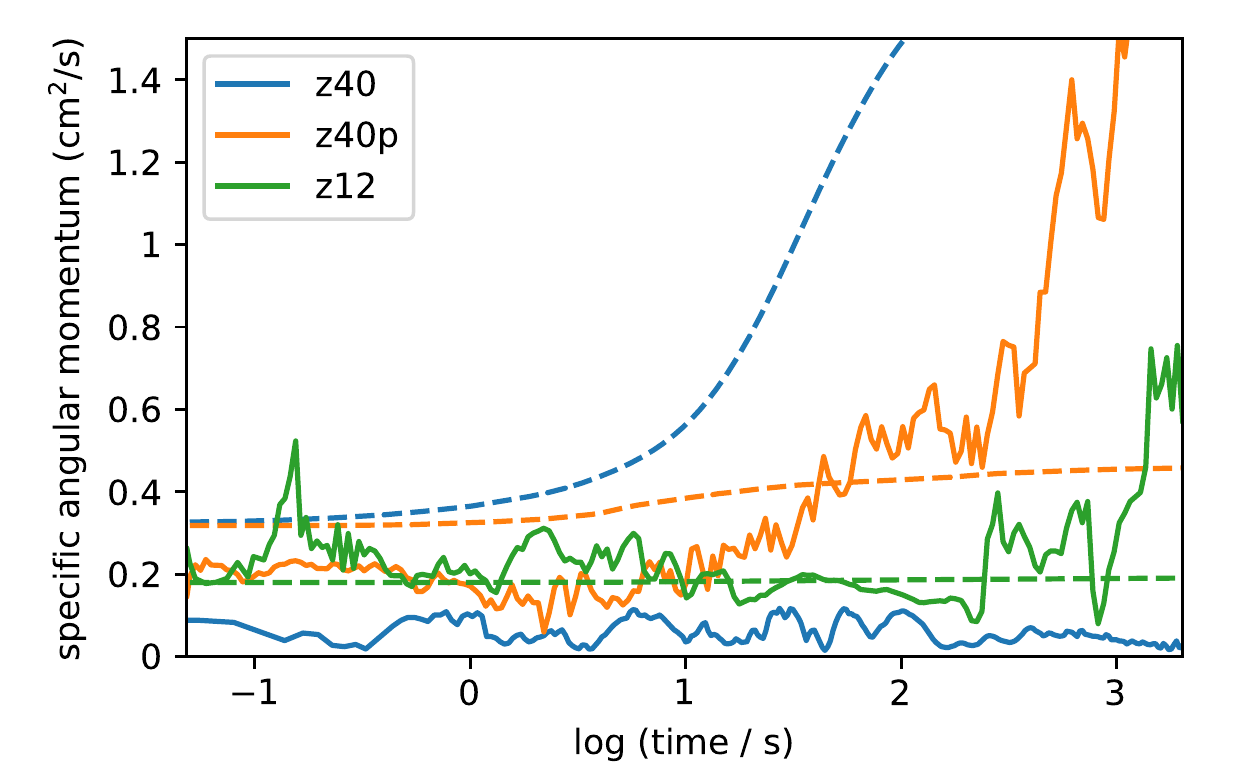}
    \caption{Specific angular momentum (\textsl{y-axis}) as a function of time after bounce (\textsl{x-axis}). \textsl{Solid lines}: specific angular momentum of the material accreting onto the central remnant. \textsl{Dashed lines}: critical angular momentum for an orbit at ISCO for $40\,\Msun$ models and at NS radius of $12\,\mathrm{km}$ for the \texttt{z12} model. Where the specific angular momentum of the infalling material exceeds this critical value, the formation of a rotationally supported disk is possible.}
    \label{fig:angmom}
\end{figure}

Our results show that late-time fallback can impart a significant kick and spin onto the central remnant, in particular \texttt{z12} exhibits an initially small angular momentum, but fallback continues to spin it up at late times to millisecond periods
(Figure~\ref{fig:bh_spin}), which may produce a magnetar-driven hypernova. Whereas accretion feedback has been considered for rotating models \citep{2019arXiv190404835B}, these effects may also be appreciable during asymmetric fallback in non-rotating explosions. At $34\,\mathrm{s}$ in \texttt{z40p}, thick transient disk-like structures can observed surrounding the central remnant in density slices (Figure \ref{fig:disk_slice}). The velocity profile of this snapshot shows that material is indeed rotating around the black hole. This is confirmed by Figure \ref{fig:angmom}, which shows that at approximately $30\,\mathrm{s}$ after bounce the instantaneous specific angular momentum of the infalling material exceeds the critical value required for a rotationally supported disk. These results are presented with several caveats. Firstly, our Newtonian approximation breaks down near the Schwarzschild radius. Secondly, whereas we have argued that our resolution is appropriate for modelling the supersonic infall of matter, it is insufficient for modelling a disk. Thirdly, we do not treat all of the physical processes that may be relevant in disks, such as radiation or magnetic fields. As such, we can only suggest that there are appropriate conditions for transient disk formation in non-rotating models, without being able to quantify any feedback into the dynamics of the explosion. We defer a detailed investigation into disk formation to a future study.

\subsection{Spin-kick alignment}
\begin{figure}
    \includegraphics[width=\columnwidth]{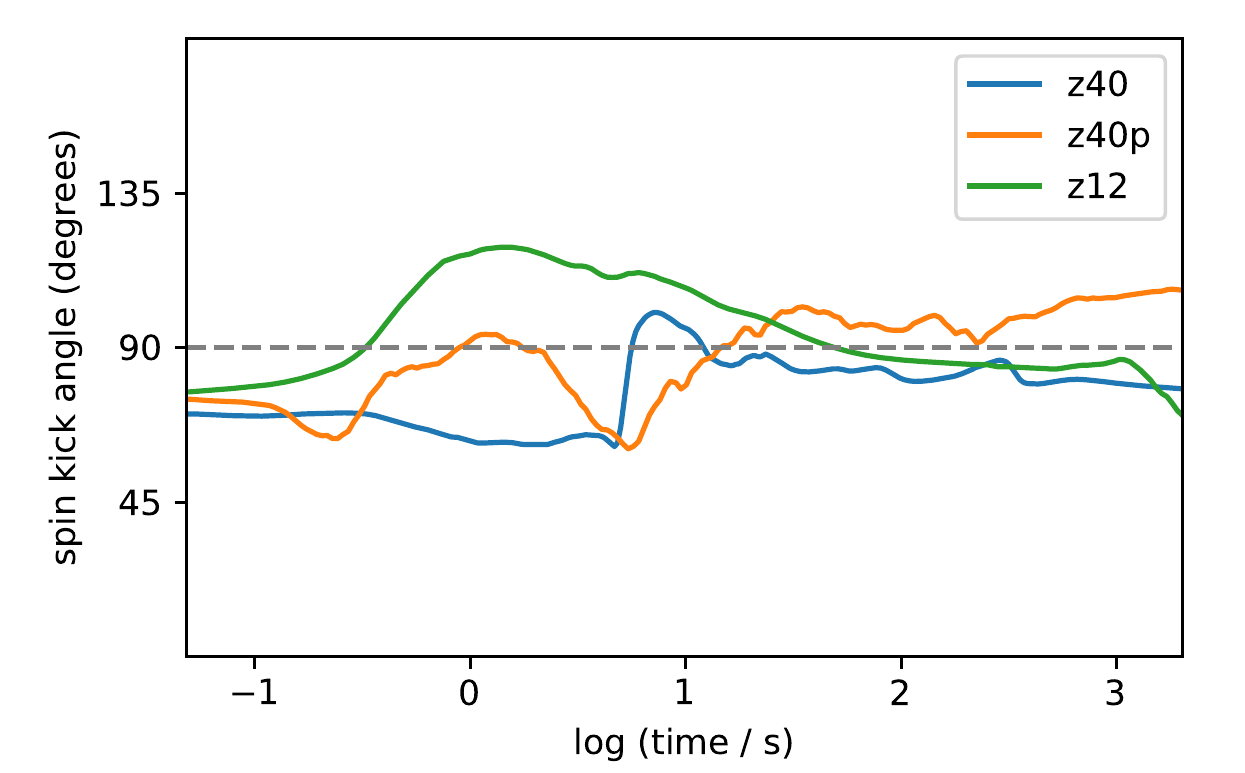}
    \caption{Angle between the kick velocity vector and spin vector of the central remnant (\textsl{y-axis}) as a function of time after bounce (\textsl{x-axis}). The dashed gray line indicates $90^\circ$ for reference.}
    \label{fig:spin_kick_alignment}
\end{figure}

Recalling that our models begin as non-rotating progenitors, we examine any potential correlations between the remnant kick and spin vectors. Radio-polarization and pulsar wind nebula observations have found evidence for  alignment between kick and spin \citep{2005MNRAS.364.1397J,2007MNRAS.381.1625J,2007ApJ...660.1357N,2012MNRAS.423.2736N}. Remnant acceleration by the emission of radiation has been suggested as an explanation for spin-kick alignment \citep{1975Natur.254..676T}. In the scenario where the kick is associated with the break-up of a binary system, it is predicted that there is a large component of the kick perpendicular to the spin vector \citep{2003pasb.conf..225C}. In the scenario where convection imparts momentum onto the remnant, the kick and spin should be uncorrelated \citep{1998Natur.393..139S}. In the event of a single momentum impulse onto the remnant which is, in general, misaligned with the remnant centre of mass, however, the spin vector is perpendicular to the kick. Considering that in this study, we quantify only the contribution to kick and spin arising from fallback of a non-rotating progenitor, we should expect to find the kick and spin are perpendicular. Unsurprisingly, we find for all three explosions that the remnant kick and spin vary at most by $30^\circ$ from perpendicular (Figure \ref{fig:spin_kick_alignment}), consistent with the accretion streams remaining relatively constant in angle. Our findings are in agreement with  numerical studies
of the gravitational tug-boat mechanism
during the first seconds of the explosion
\citep{2013A&A...552A.126W,2017MNRAS.472..491M,2019MNRAS.484.3307M}, and suggest that fallback is not responsible for any observed spin-kick alignments.

\subsection{Chemical yields}
The mixing-fallback scenario allows the chemical yields of stellar models to be tuned to explain the low abundance of iron despite the presence of lighter elements in present-day metal-poor stars \citep{2002ApJ...565..385U,2003Natur.422..871U,2005ApJ...619..427U,2014Natur.506..463K}.  We investigate the plausibility of this process using our models. As shown in \cite{2018ApJ...852L..19C}, \texttt{z40} fails to eject \emph{any} elements other than the hydrogen envelope and does not explain these observations, thus we look to the more energetic \texttt{z40p} for evidence of more mixing and ejection. Our models fall short of a detailed evaluation of abundance yields for several reasons. In the \textsc{CoCoNuT-FMT} models, the freeze-out from NSE is approximated using a threshold temperature with a small NSE composition table, and a ``flashing'' prescription for the burning of selected $\alpha$-elements is used below the NSE temperature. In addition, there are uncertainties in the electron fraction as discussed by \citet{2017MNRAS.472..491M,2018MNRAS.479.3675M}. We also cease the treatment of nuclear reactions after mapping to to \textsc{Arepo}, on the assumption that further contributions to the \emph{dynamics}, rather than the nucleosynthesis, are insignificant. Nonetheless, we can adequately quantify the effects of the mixing and fallback process on the final yield. Given the uncertainties in the nucleosynthesis, it is more meaningful to show the total iron-group composition, rather than the individual constituent isotopes.

\begin{figure}
    \includegraphics[width=\columnwidth]{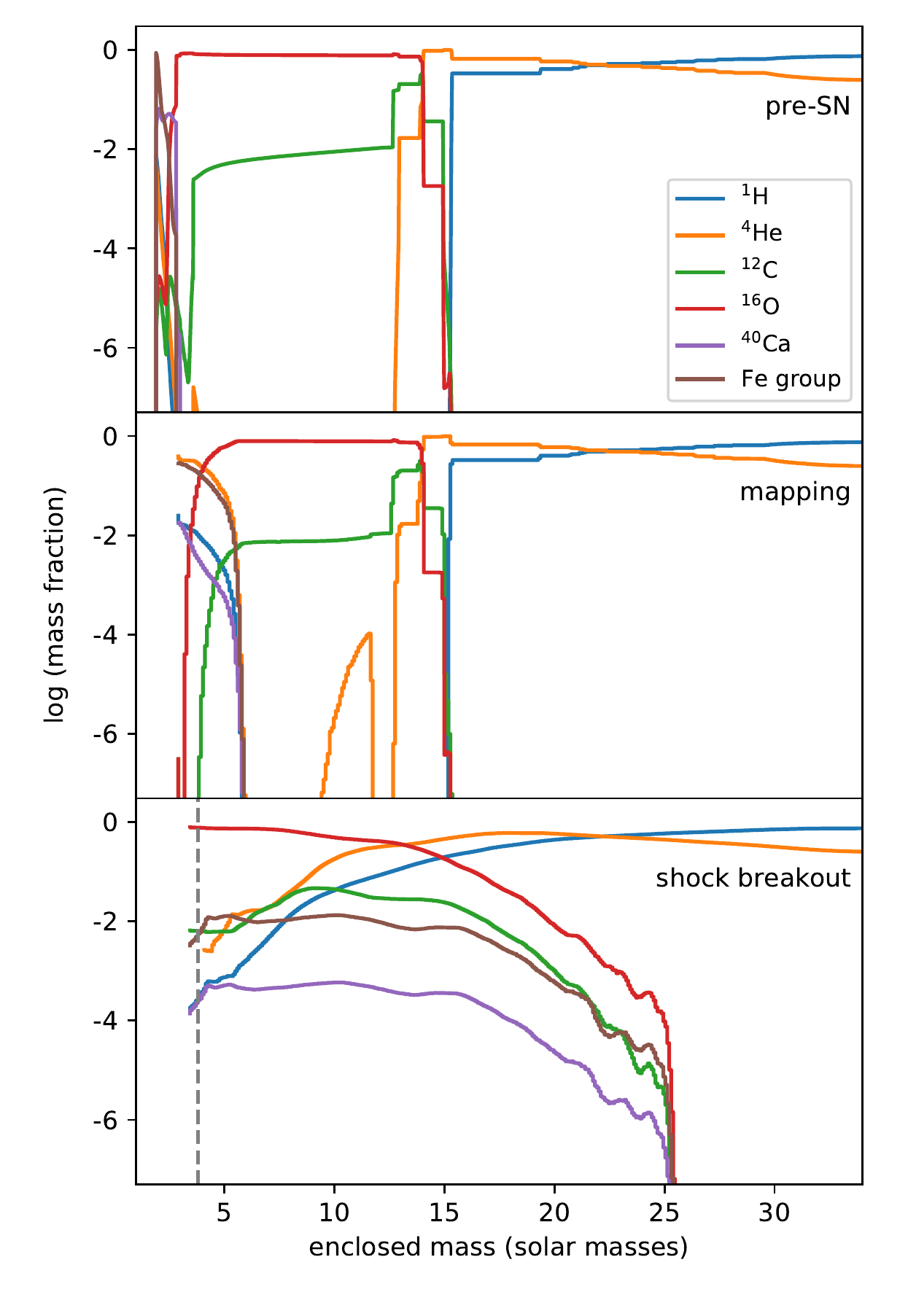}
    \includegraphics[width=\columnwidth]{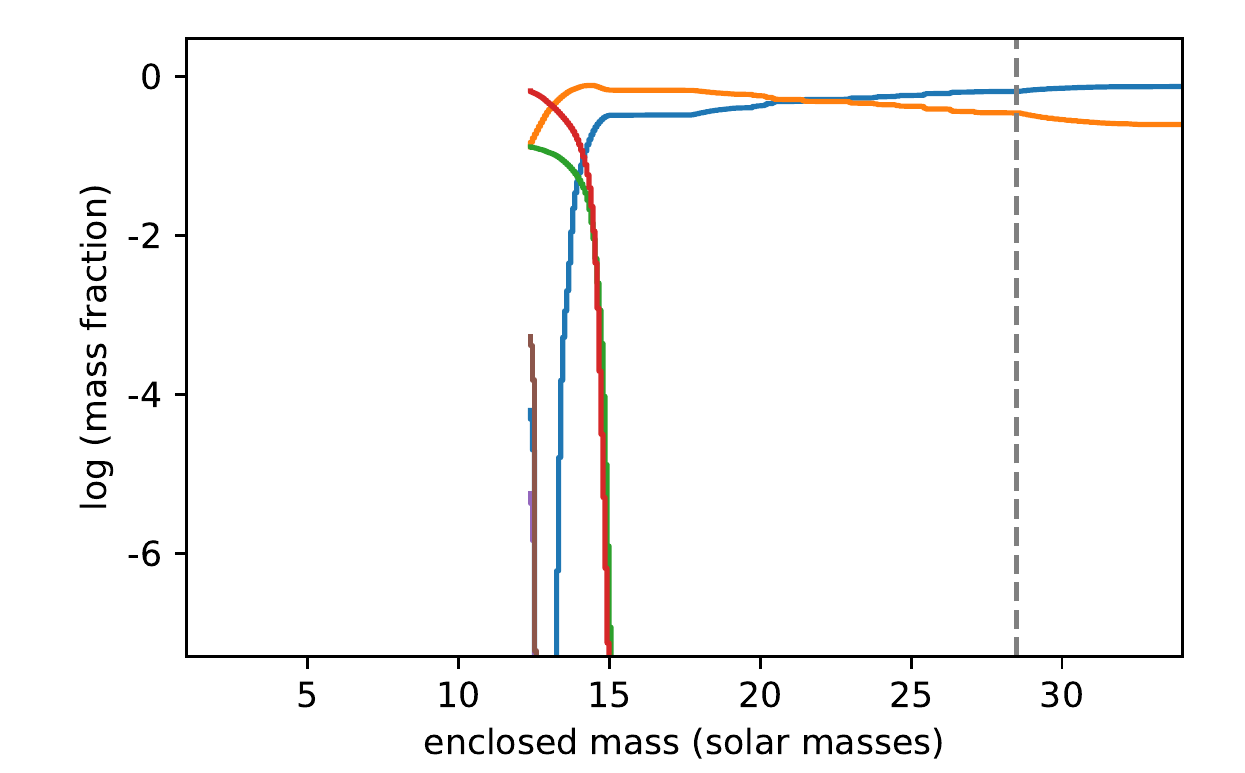}
    \caption{\textsl{Upper Panel:} Distribution of elements as a function of mass coordinate prior to collapse, at the time of BH formation, and at shock breakout for \texttt{z40p}. The \textsl{gray dashed line} indicates the predicted mass cut.  \textsl{Lower Panel:} The distribution of elements at shock breakout for \texttt{z40}, showing that only the envelope is ejected.}
    \label{fig:xm_chem}
\end{figure}

\begin{figure*}
    \includegraphics[width=\textwidth]{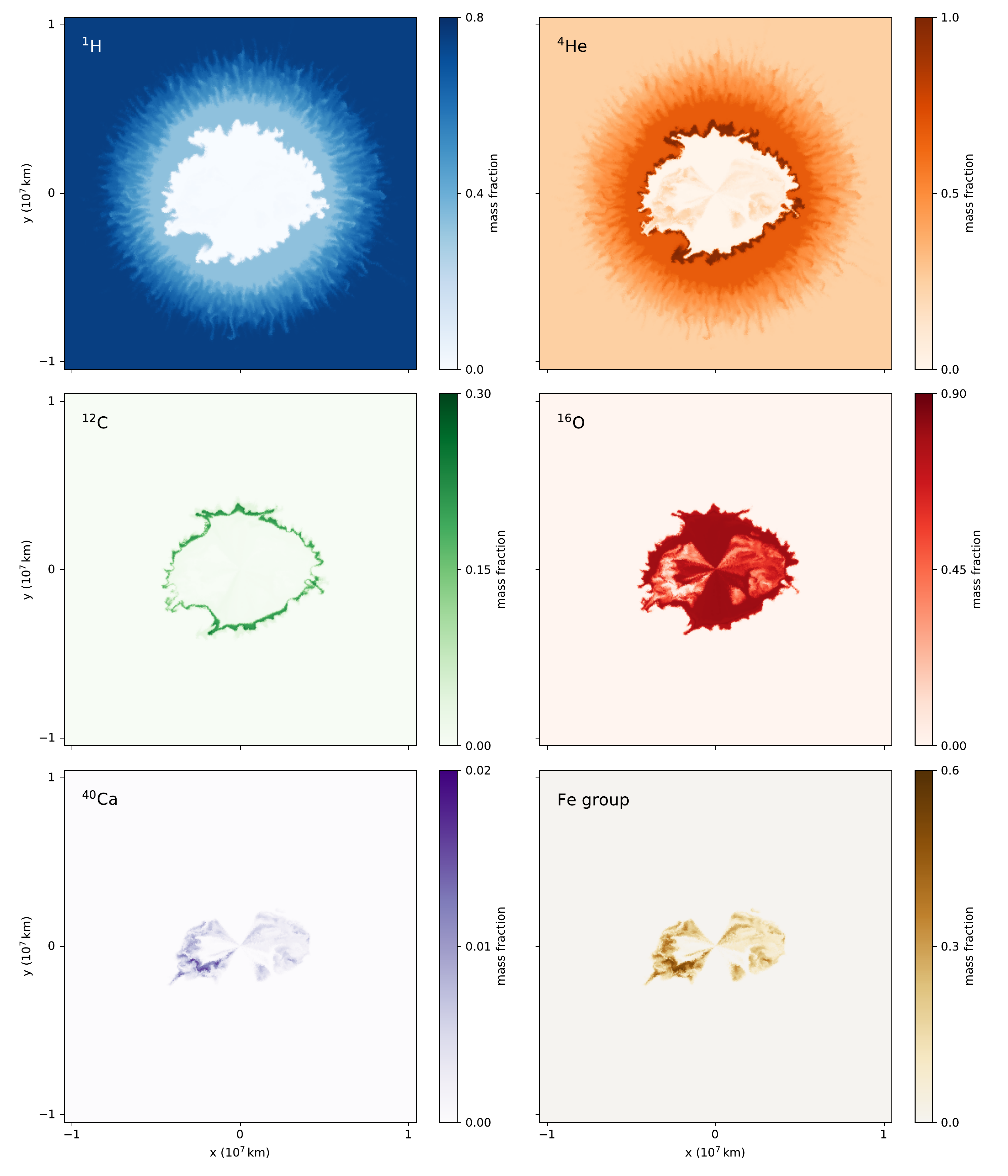}
    \caption{Slices showing mass fraction of elements at shock breakout for \texttt{z40p}.  Colours correspond to species as in Figure~\ref{fig:xm_chem}.}
    \label{fig:2d_chem}
\end{figure*}

Figure \ref{fig:xm_chem} shows the spherically averaged distribution of elements at the time of shock breakout. The plumes shown in Figure \ref{fig:2d_chem} indicate mixing arising from the Rayleigh-Taylor instability, a phenomenon well documented by multi-dimensional mixing studies \citep{1991A&A...251..505M,1991ApJ...367..619F,1990ApJ...358L..57H,2003A&A...408..621K,2009ApJ...693.1780J,2010ApJ...723..353J,2015A&A...577A..48W,2017MNRAS.467.4731C}. Our models show only limited mixing by
the Rayleigh-Taylor instability at
the H/He interface, in agreement with
other simulations of mixing in Pop~III
stars \citep{2009ApJ...693.1780J} that were
based on piston-driven explosion model.
This is due to the more compact envelope
of Pop~III stars. In our models, carbon, oxygen, calcium, and iron group elements are well mixed up to the helium shell
in terms of their radial distribution, and weakly into the envelope. In terms of their 3D distribution, these elements retain
a distinctly different distribution, though.
All of this material, however, is well above the estimated mass cut of $3.82\,\Msun$, and will be ejected. Due to the large-scale asymmetric flows, which push material out from the core and break the layered structure in composition, we do not see any evidence for selective trapping of the iron-group elements. In this model,
fallback cannot efficiently separate
iron-group elements and intermediate
mass elements.

What we find are thus two extreme scenarios, one with layered
ejection and no metal ejecta (\texttt{z40}), and one with over-abundantly iron-rich ejecta (\texttt{z40p}). Neither of those
scenarios matches the abundance patterns with a low ratio
of iron-group to intermediate-mass element
abundances as found in some extremely metal-poor stars \citep{2002Natur.419..904C, 2004A&A...416.1117C,2005Natur.434..871F,2014Natur.506..463K,2015ApJ...806L..16B}.
Though the possibility of explaining the abundances of metal-poor stars using fallback has not yet been ruled out, it is still contingent on identifying a successful explosion mechanism that preserves the layered structure of the star.
It is worth recalling that \texttt{z40p} has an artificially enhanced explosion energy, and an explosion with energy between the two might reproduce the
required yields.

\begin{figure}
    \includegraphics[width=\columnwidth]{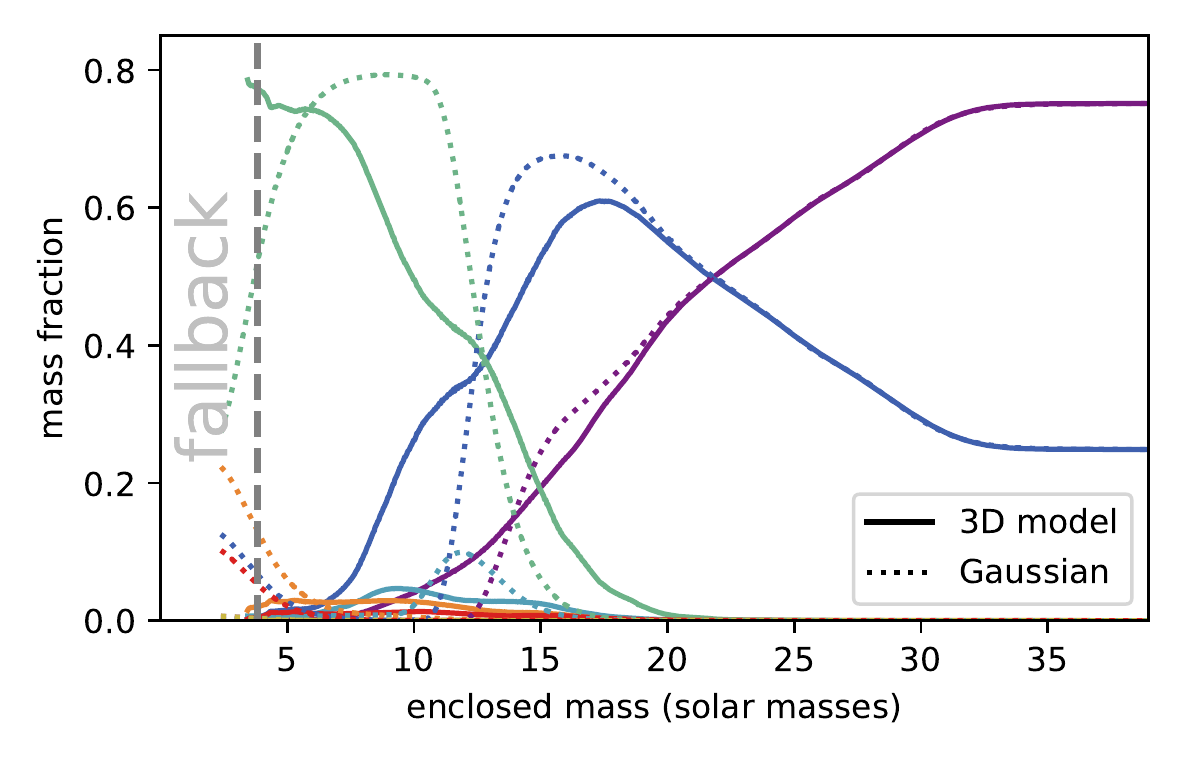}
    \includegraphics[width=\columnwidth]{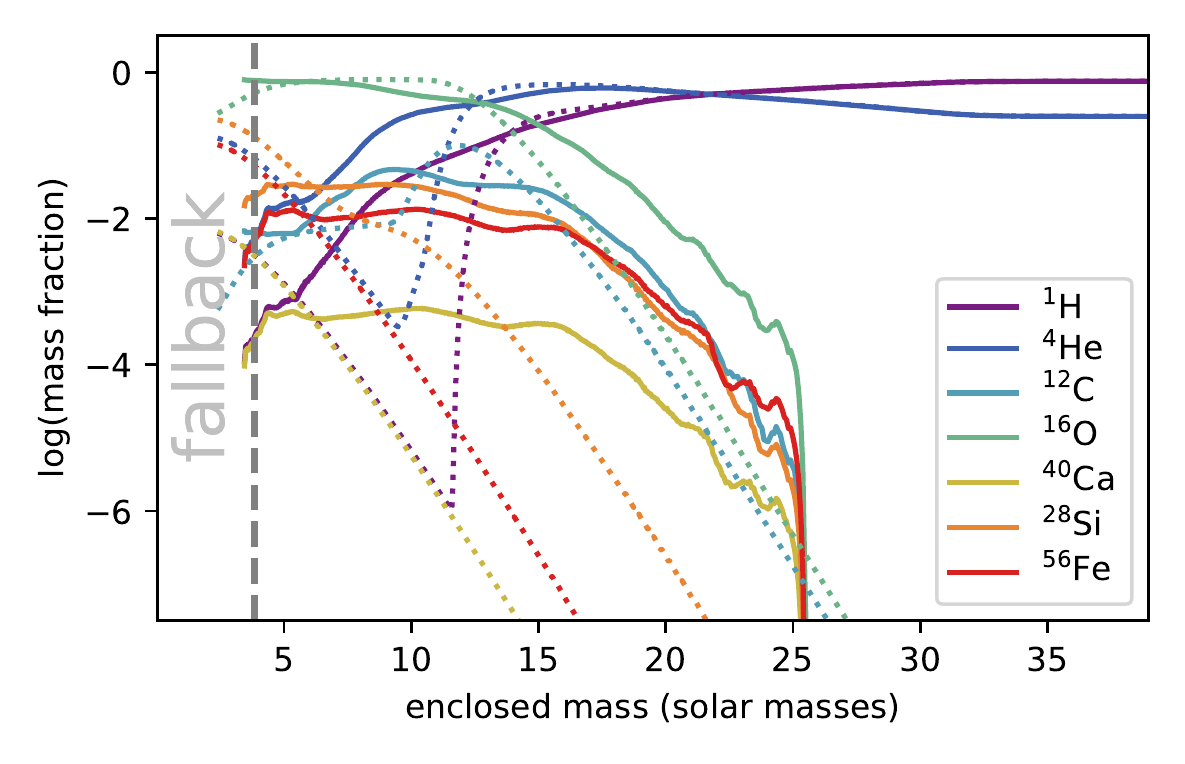}
    \caption{Distribution of elements as a function of mass coordinate at shock breakout for \texttt{z40p} on linear (\textsl{Upper Panel}) and logarithmic (\textsl{Lower Panel}) scales.  The \textsl{gray dashed line} indicates the predicted mass cut.  The \textsl{dotted lines} indicate the distribution based on a Gaussian smoothing kernel that best matches the total yield of the full model.  The broad mixing before a sharp drop-off cannot be reproduced.}
    \label{fig:mix}
\end{figure}

Unlike SN light curves and spectra, SN yields used in galactic chemical evolution (GCE) studies only depend on the total yield, not the spatial distribution of elements within the ejecta. In GCE studies, ``sub-grid'' modelling of SN ejecta composition using efficient parameterised mixing approximations allows for the yields of thousands of progenitor models to be realised.  We analyse the accuracy of one such approximation, diffusive mixing, by comparing the distribution of elements at shock breakout with the diffusive mixing model that provides the closest \emph{total} yield.  This resulting mixing is similar to simpler mixing algorithms (e.g., boxcar filters) that have been used in the past \citep[e.g.,][]{1995ApJS..101..181W,2002ApJ...576..323R,2010ApJ...724..341H} and that was successful in reproducing the light curve of SN~1987A at the time \citep{1988ApJ...330..218W}.  In Figure~\ref{fig:mix} we show the results of mass-coordinate Gaussian smoothing with a width of $0.0436$ times the mass of the helium core ($15.29\,\Msun$).  This corresponds to approximately a mixing parameter of $0.025$ for the boxcar mixing model used by \citet{2010ApJ...724..341H}, consistent with low mixing as suggested by \citet{2009ApJ...693.1780J} for primordial stars, and to what was found as best observational match by \citet{2015ApJ...806L..16B}.  Specifically, we solve for each species the implicit equation 
\begin{equation}
X_i^+ - X_i^- = -\Delta\!M^2\,\frac2{m_i}\left(
\frac{X_{i+1}^+-X_i^+}{m_{i+1}+m_i}-\frac{X_{i}^+-X_{i-1}^+}{m_{i}+m_{i-1}}\right)\;,
\end{equation}
where $\Delta\!M$ is the constant mixing mass, $m_i$ and $X_i$ are the mass and mass fraction of zone $i$, and the superscripts ``$^-$'' and ``$^+$'' indicate the initial and mixed abundances, and we use reflective boundaries.

The lower panel of Figure~\ref{fig:mix} shows that despite reproducing the total yield, the distribution of species toward the envelope cannot be well reproduced by this mixing model. Parameterised mixing-fallback models employed to reproduce the star of \citet{2014Natur.506..463K} and other carbon-enhanced metal-poor (CEMP) stars usually require a mass cut around the edge of the helium core or the edge of the CO core. This is where we see significant impact of the full 3D modelling. For example, in the case shown, the $^{40}$Ca to $^{16}$O yield outside the helium core (mass coordinate $15\,\Msun$) is vastly different between the 3D simulation and the Gaussian mixing model.  Some more sophisticated descriptions for supernova mixing have recently been suggested \citep[e.g.,][]{2016ApJ...821...76D}, but how well those compare would require more detailed studies that are beyond the scope of this paper.  We conclude that one has to be careful when using ad-hoc prescriptions in the mixing and fallback model for obtaining yields for matching with observed CEMP star abundance patterns.

Whereas \texttt{z40} and \texttt{z40p} have too much and too little fallback, respectively, to reproduce the
extremely iron-poor star SMSS J031300.36-670839.3
\citep{2014Natur.506..463K}, it is reasonable to expect that some intermediate explosion energy and realisation of asymmetries may well be able to reproduce the star's abundance pattern.  It is rather interesting to note, though, that in \texttt{z40p} a large amount of $^{56}$Fe, $^{28}$Si, and $^{40}$Ca is mixed out well beyond the edge of the helium core, and if that was ejected, it would be clearly inconsistent with \citep{2014Natur.506..463K}, which is known for its low [Ca/H] and and upper limit on [Fe/H] of around -7.

\section{Conclusions}
We have presented supernova mixing-fallback models calculated using \textsc{CoCoNuT-FMT} code \citep{2015MNRAS.448.2141M} followed by the \textsc{Arepo} code \citep{2010MNRAS.401..791S} from 
$12\,\Msun$ and $40\,\Msun$ 
Pop~III models of \cite{2010ApJ...724..341H}.  We have shown that the effects of fallback could be sufficient to reproduce substantial kicks and spins inferred for black holes and neutron stars for a suitable ratio of the initial energy
and envelope binding energy.
Strongly asymmetric fallback happens if
the ratio is low enough for substantial fallback to occur, but high enough to
avoid complete fallback of the asymmetric inner ejecta.

In our models, fallback can still change the remnant kicks and spins on timescales of hundreds to thousands of seconds.  As an extreme case, the $12\,\Msun$ explosion model shows that late time fallback may spin up the nascent neutron star to millisecond periods, provided that there is no mechanism \citep[e.g.,][]{2011ApJ...736..108P} to stop the  accretion of angular momentum. 
Our results suggest that there should be a positive correlation between spin and kick magnitudes, and that fallback leads to produce misaligned spins and kicks. If at all, there seems to be  preference for perpendicular spins and kicks.  As a consequence of momentum conservation, we posit that large black holes may have low spins and kicks, whereas small black holes could have large spins and kicks.

Our models also hint at the possibility of disk formation, which may slow the remnant accretion rate, but also drive outflows that can result in observable transients \citep{2013ApJ...772...30D, 2018ApJ...867..130F,2019ApJ...880...21M}.  Disk formation around the central remnant will have to be modelled using a higher resolution simulation with treatment for radiation and magnetic fields to quantify the feedback effects and identify any transients that may be produced. 

In the more energetic explosion of the $40\,\Msun$ star, we found that elements heavier than carbon were thoroughly mixed in the core during the explosion, so that fallback did not separate the iron from being ejected. Although the abundance ratios in ejecta fail to explain the iron-poor abundance of CEMP stars, this model provides a valuable scenario in the upper range of ejected iron content, and suggests that an explosion energy within the two energies we simulated may produce a compatible layered fallback abundance pattern. The contrast between our 3D calculations and the 1D diffusion approximation, which better preserve the layered structure in composition, is unsurprising given that 1D approximations cannot self-consistently resolve large-scale asymmetries in the fluid flow.

In this study we have only just begun examining whether the ideas about fallback traditionally used to interpret observations are actually borne out in nature. It is evident from our selection of models that fallback can have a substantial impact on the properties of compact remnants and chemical ejecta produced by CCSNe. These reassuring prospects motivate further simulations to explore the full spectrum of fallback scenarios.

\section*{Acknowledgements}

We thank Volker Springel and R\"udiger Pakmor for the help they provided with the \textsc{Arepo} code. This research was undertaken with the assistance of resources and services from the National Computational Infrastructure (NCI), which is supported by the Australian Government.  It was supported by resources provided by the Pawsey Supercomputing Centre with funding from the Australian Government and the Government of Western Australia. CC  was supported by an Australian Government Research Training Program (RTP) Scholarship. AH and BM were supported by ARC Future Fellowships FT120100363 (AH) and FT160100035 (BM). This work was supported by JINA-CEE through US NSF grant PHY-1430152. BM was supported by STFC grant ST/P000312/1.
Parts of this research were conducted by the Australian Research Council Centre of Excellence for Gravitational Wave Discovery (OzGrav), through project number CE170100004.  
AH has been supported, in part, by the Australian Research Council Centre of Excellence for All Sky Astrophysics in 3 Dimensions (ASTRO 3D), through project number CE170100013; and by a grant from Science and Technology Commission of Shanghai Municipality (Grants No.16DZ2260200) and National Natural Science Foundation of China (Grants No.11655002).  

\bibliographystyle{mnras}
\bibliography{library}

\bsp
\label{lastpage}
\end{document}